\documentstyle[12pt]{article}
\newcommand{\beq}{\begin{equation}}
\newcommand{\eeq}{\end{equation}}
\newcommand{\beqn}{\begin{eqnarray}}
\newcommand{\eeqn}{\end{eqnarray}}
\def\vdir{v\kern-7.8pt\Big{/}}
\def\pdir{p\kern-7.8pt\Big{/}}

\begin{document}

\title{Cancellation of power enhancements in non-spectator decays}
\vskip 2.5truecm
\author{
U.~Aglietti\\
SISSA-ISAS, Via Beirut 2, 34014 Trieste, Italy \\
INFN, Sezione di Trieste, Via Valerio 2, 34100 Trieste, Italy\\}
\date{}
\maketitle
\begin{abstract}
\noindent

Exclusive non-spectator decay rates of beauty hadrons
contain power enhancements of the form $(m_b/m)^2$ and $m_b/m$,
where $m_b$ is the $b$-quark mass and $m$ is a light quark mass.
An implicit argument has been recently given according
to which these singularities cancel in the totally inclusive decay width.
We present in this note a completely explicit computation of the diagrams
containing power enhancements. Our results agree with the previous
general conclusion.

\end{abstract}

\newpage

The basic decay mechanism of beauty hadrons is the decay
of the constituent $b$-quark.
The spectator model assumes the dominance of this mechanism,
neglects binding corrections of the $b$-quark
in the hadron, and
computes inclusive decay widths by means of the parton model.
For example, the semileptonic decay
$
H_b~\rightarrow~X_c+l+\nu,
$
where $H_b$ is a beauty hadron and
$X_c$ is any hadronic final state with charm $C=1$, is computed as
the decay
$
b\rightarrow c+l+\nu.
$
The rate is given by:
\beq\label{eq:semil}
\Gamma~=~\Gamma_{0}[1-(2\alpha_s/3\pi)f(m_c/m_b)]~I(m_c/m_b,m_l/m_b,0)
\eeq
where
$\Gamma_0=G_{F}^2m_b^5\mid V_{cb}\mid^2/192\pi^3$,
$f(x)$ is a perturbative $QCD$ correction with $f(0)=\pi^2-25/4$, and
$I(x_1,x_2,x_3)$ is a phase-space correction factor for non-zero masses
in the final state, with $I(0,0,0)=1$ \cite{cortes}.

\noindent
Hadronic decays are computed in a quite analogous way. The rate for the decay
\beq\label{eq:hadch}
b~\rightarrow~ c+\overline{u}+d,
\eeq
for example, is given by:
\beq\label{eq:hadr}
\Gamma=3\Gamma_{0}\mid V_{ud}\mid ^2
\frac{2C_{+}^2+C_{-}^2}{3}J I(m_c/m_b,m_d/m_b,m_u/m_b)
\eeq
where $C_{+,-}=(\alpha_s(m_b)/\alpha_s(m_W))^{d+,-}$,
with $d_{+,-}=-6/23,~12/23$, are coefficients
resumming leading-logs coming from the renormalization of the non-leptonic
hamiltonian \cite{altmai}
(numerically, $C_{+}\simeq 0.88$ and $C_{-}\simeq 1.29$),
and $J$ is a factor including $QCD$-subleading corrections
computed in ref.\cite{sublead}.

\noindent
An important consequence of the spectator model is that
all beauty hadrons should have the same lifetime, because the light quark(s)
in the initial state behave like passive spectators.
$LEP$ experiments aim at testing this prediction with errors less
than the expected splitting.
The semileptonic branching fraction $B_{SL}$ can also be predicted with
reasonable accuracy.
This is because the leading dependence on $m_b^5$ ($m_b$
is not clearly defined), cancels in taking the ratio of widths.
A detailed computation including also penguin diagrams,
gives a semileptonic branching-ratio in the range \cite{altpetr}:
$
B_{SL}~=~[12.2\pm 0.45(scale)\pm 0.8(\alpha_s)]\%,
$
for current values of quark masses, and
$
B_{SL}~=~[14.4\pm 0.45(scale)\pm 0.8(\alpha_s)]\%,
$
for constituent values of quark masses.
The first uncertainty comes from the scale of the coupling $\alpha_s$,
while the second one comes from the experimental error on $\alpha_s(m_Z)$.
The experimental value
$
B_{SL}~=~(10.5\pm 0.5)\%
$\cite{pp}
is marginally consistent with the prediction of the spectator model,
implying non-spectator effects of the order of $15\%$.

There is a clear indication of the relevance of
decay mechanisms which lie outside the spectator model,
involving the light quark(s) in the hadron in an essential way
(non-spectator mechanisms).

Non-spectator mechanisms are of various kind and involve different
quark-level processes.
The corresponding rates are proportional to $m_b^n$ with $n\leq 3$.
Since the partonic rate is proportional to $m_b^5$, all non-spectator effects
are preasymptotic in the limit of a very large mass of the heavy quark,
$m_b\rightarrow\infty$.

Let us recall very briefly the corrections related to the (not so free)
$b$-decay.
Partonic rates, such as those in eqs.(\ref{eq:semil}) and (\ref{eq:hadr}),
have corrections for the binding of
the $b$-quark in the hadron (bound-state corrections \cite{accmm}),
as well as non-perturbative corrections for final state interactions.
Recently there has been some progress
in the model-independent computation of this kind of corrections
using the heavy quark effective theory ($HQET$) \cite{hqet}.
The semileptonic branching fraction of $B$ mesons is modified
according to \cite{nonper}:
\beq\label{eq:nonper}
\frac{\delta B_{SL}(B)}{B_{SL}(B)}
{}~\approx~6\frac{M_{B*}^2-M_B^2}{m_b^2}\frac{C_+^2-C_-^2}{2N}~\approx~-2\%
\eeq
$B_{SL}$ therefore is reduced with respect to the
spectator value in the 'right direction',
although by a tiny amount only. The effect is small because
color dynamics gives a large suppression factor:
$(C_+^2-C_-^2)/2N\simeq -0.15$.
$\delta B_{SL}/B_{SL}$ vanishes instead in the $\Lambda_b$ case.
As it stems from eq.(\ref{eq:nonper}),
these corrections are of order $1/m_b^2$ with respect to the partonic rate.

We now consider corrections related to different quark-level processes
with respect to $b$ decay.
The case of $B$-mesons is discussed for simplicity.
The neutral $B_d$ meson can decay via annihilation of the constituents $b$ and
$\overline{d}$ quarks \cite{bss}.
The corresponding decay rate is given by:
\beq\label{eq:ann}
\frac{ \delta\Gamma_{ann} }{\Gamma_0}~=~
24\pi^2\left(\frac{f_B}{m_b}\right)^2
\left(\frac{m_c}{m_b}\right)^2\left[
3\left(\frac{C_+-C_-}{2}\right)^2+\frac{C_+^2+C_-^2}{2}
+\frac{1}{3}\left(\frac{C_++C_-}{2}\right)^2
\right]
\eeq
where $f_B$ is the decay constant of the $B$ meson
(numerically, $f_B\sim 200~MeV$ from lattice $QCD$ computations \cite{allt}).
The width in eq.(\ref{eq:ann}) is very small,
$\delta\Gamma_{ann}/\Gamma_0\sim 5\%$,
because there is a large helicity-suppression factor $(m_c/m_b)^2\sim 0.1$.
The annihilation mechanism increases the decay rate of the $B^0$ meson with
respect to the spectator value.

The decay of charged $B_u$ mesons is modified with respect to the
spectator case by an interference effect, because
in the channel (\ref{eq:hadch}) there are
two identical $\overline{u}$-quarks in the final state \cite{interf, interf2}.
The modification of the decay rate due to the interference is given by:
\beq\label{eq:inter}
\frac{ \delta\Gamma_{int} }{\Gamma_0}~=~
48\pi^2\left(\frac{f_B}{m_b}\right)^2\left[
             \frac{C_+^2-C_-^2}{2}+\frac{C_+^2+C_-^2}{2N}\right]
\eeq
The rate in eq.(\ref{eq:inter}) is negative, i.e. the
interference reduces the decay rate of the charged $B$'s with respect to the
partonic value. In this case, there is no more the helicity-suppression
factor present in eq.(\ref{eq:ann}), but the non-spectator rate is again
very small because of color dynamics.
There is a large cancellation between the two terms in
the square bracket in eq.(\ref{eq:inter}), which
is efficient up to $\sim 90\%$, and gives
$\delta\Gamma_{int}/\Gamma_0\sim -3\%$.

As it stems from eqs.(\ref{eq:ann}) and (\ref{eq:inter}),
the annihilation and interference rates are
proportional to $m_b^2$, because the decay constant scales as
$f_M\propto 1/\sqrt{m_b}$ for very large masses \cite{pwsv}.
These mechanisms therefore are subleading with respect to the
non-perturbative corrections discussed before.
As a consequence,
we expect for very large $m_b$ a greater difference
in lifetimes between mesons and
baryons, than the ones inside mesons.

Radiative corrections to non-spectator decays may in principle increase
considerably the rates. In the case of the interference, for example,
gluon radiation from the constituent light quark contributes to the decay
rate by a term of the form:
\beq\label{eq:int_gl}
\frac{ \delta\Gamma_{int}' }{\Gamma_0}~=~
\pi\alpha_s\left(\frac{f_B}{m}\right)^2(1-1/N^2)\frac{C_+^2+C_-^2}{2}
\eeq
There is a power enhancement of the form $(m_b/m)^2$
with respect to the lowest order term, which overcomes the typical
suppression factor $\alpha_s/4\pi$ of the radiative corrections.
Furthermore, color dynamics is different from that one
of the tree amplitude, and
the almost complete cancellation between $C_+$ and $C_-$ terms
in eq.(\ref{eq:inter}) does not occur any more.
Formula (\ref{eq:int_gl}) implies that non-spectator effects may be
substantially greater than $1\%$ level as implied by the tree-level
computations (\ref{eq:ann}) and (\ref{eq:inter}).
However, the rate in eq.(\ref{eq:int_gl})
depends critically on the value of the light quark mass $m$,
which cannot be reliably estimated since it involves unknown non-perturbative
dynamics. The result therefore has a limited phenomenological
impact. Analogous power enhancements occur in the radiative corrections
to the annihilation width of $B^0$ mesons.

In ref.\cite{centrale} a brilliant
analyticity argument has been given according to which all
power enhancements cancel in the totally inclusive decay rate, leaving
only residual logarithmic singularities of the form $\log(m_b/m)$.

In this note we present an explicit computation of the diagrams
containing power enhancements. For definiteness we consider the
interference case.
The absorptive part of the diagram in fig.1 is given by the sum of the
cuts $(i),~(ii)$ and $(iii)$:
\beqn
{\cal M}&=&{\cal M}_{(i)}+{\cal M}_{(ii)}+{\cal M}_{(iii)}=
C\int\int\frac{d^4k}{(2\pi)^4}\frac{d^4l}{(2\pi)^4}
\overline{v}_u(p_2)ig\gamma_{\sigma}
\\ \nonumber
&&i\frac{-\hat{k}-\hat{p}_2+m}{(k-p_2)^2-m^2+i\epsilon}
(-i)\frac{G_F}{\sqrt{2}}\gamma_{\mu}(1-\gamma_5)(\hat{P}-\hat{k}-\hat{l})
2\pi\delta^+((P-k-l)^2)
\\ \nonumber
&&(-i)\frac{G_F}{\sqrt{2}}\gamma_{\nu}(1-\gamma_5)
i\frac{-(\hat{k}+\hat{p}_4)+m'}{(k+p_4)^2-{m'}^2+i\epsilon}
ig\gamma_{\rho}v_u(p_4)
\\ \nonumber
&&\frac{-ig^{\rho\sigma}}{k^2-\lambda^2+i\epsilon}
\overline{u}_b(p_3)\gamma^{\mu}(1-\gamma_5)\hat{l}2\pi\delta^+(l^2)
\gamma^{\nu}(1-\gamma_5)u_b(p_1)
\\ \nonumber
&&(-2\pi i)~\{~((k+p_4)^2-{m'}^2)\delta^+((k+p_4)^2-{m'}^2)+
\\ \nonumber
&&(k^2-\lambda^2)\delta^+(k^2-\lambda^2)+
((k+p_2)^2-m^2)\delta^+((k+p_2)^2-m^2)         \}
\eeqn
The gluon propagator is taken for simplicity in the Feynman gauge
and infrared divergences are regulated with a fictious gluon mass $\lambda$.
$P=p_1-p_4$, $p_1$ and $p_2$ are the initial momenta of the $b$ and
$\overline{u}$ quark respectively, and $p_3$ and $p_4$ are the final momenta.
The masses of the $c$ and the $d$ quarks have been taken equal to zero
since they do not give rise to mass singularities
(see later for the definition of $m$ and $m'$).
$C=C^{ik}_{je}=
1/2(C_+^2+C_-^2)(1-1/N^2)\delta^{i}_{j}\delta^{k}_{e}$ is a color factor.

Computing the spin algebra, we see that the amplitude ${\cal M}$ can be
decomposed according to the powers of the masses $m$ and $m'$ appearing at
the numerator:
\beq
{\cal M}~=~{\cal M}^{(0)}+m'{\cal M}^{(1a)}+m{\cal M}^{(1b)}+mm'{\cal M}^{(2)}
\eeq
The loop over $l$ is easily integrated and we have:
\beqn
{\cal M}^{(0)}&=&-\frac{\alpha_s G_F^2C}{2\pi^3}
\overline{v}_u(p_2)\gamma^{\nu}(1-\gamma_5)v_u(p_4)
\overline{u}_b(p_3)\gamma^{\mu}(1-\gamma_5)u_b(p_1)I_{\mu\nu}~~~~~~~
\\
{\cal M}^{(1a)}&=&-\frac{\alpha_s G_F^2C}{\pi^3}
\overline{v}_u(p_2)(1+\gamma_5)v_u(p_4)
\overline{u}_b(p_3)\gamma^{\mu}(1-\gamma_5)u_b(p_1)V_{\mu}
\\
{\cal M}^{(1b)}&=&-\frac{\alpha_s G_F^2C}{\pi^3}
\overline{v}_u(p_2)(1-\gamma_5)v_u(p_4)
\overline{u}_b(p_3)\gamma^{\mu}(1-\gamma_5)u_b(p_1)W_{\mu}
\\
{\cal M}^{(2)}&=&-\frac{\alpha_s G_F^2C}{2\pi^3}
\overline{v}_u(p_2)\gamma^{\mu}(1+\gamma_5)v_u(p_4)
\overline{u}_b(p_3)\gamma_{\mu}(1-\gamma_5)u_b(p_1)S
\eeqn
where $I_{\mu\nu}=I_{\mu\nu}^{(i)}+I_{\mu\nu}^{(ii)}+I_{\mu\nu}^{(iii)}$
and similarly for the other integrals.
For the cut $(iii)$, we have:
\beqn
I_{\mu\nu}^{(iii)}&=&\int\frac{ d^4k~\delta^+((k+p_2)^2-m^2) }
       {  (k^2-\lambda^2+i\epsilon)((k+p_4)^2-{m'}^2+i\epsilon)  }
\{(k+p_2)_{\mu}(k+p_4)_{\nu}~~~
\\ \nonumber
&&~~~+(k+p_2)_{\nu}(k+p_4)_{\mu}
-g_{\mu\nu}(k+p_2)(k+p_4)
\\ \nonumber
&&~~~-i\epsilon_{\rho\sigma\mu\nu}(k+p_2)^{\rho}(k+p_4)^{\sigma}~\}
{}~\frac{1}{2}(P-k)^2\Theta ^+((P-k)^2)
\\ \nonumber
V_{\mu}^{(iii)}&=&\int\frac{ d^4k~\delta^+((k+p_2)^2-m^2) }
       {  (k^2-\lambda^2+i\epsilon)((k+p_4)^2-{m'}^2+i\epsilon)  }
          (k+p_2)_{\mu}\times
\\ \nonumber
&&~~~~~~~~~~\times\frac{1}{2}(P-k)^2\Theta ^+((P-k)^2)
\\ \nonumber
W_{\mu}^{(iii)}&=&\int\frac{ d^4k~\delta^+((k+p_2)^2-m^2) }
       {  (k^2-\lambda^2+i\epsilon)((k+p_4)^2-{m'}^2+i\epsilon)  }
(k+p_4)_{\mu}\times
\\ \nonumber
&&~~~~~~~~~~\times\frac{1}{2}(P-k)^2\Theta ^+((P-k)^2)
\\ \nonumber
S^{(iii)}&=&\int\frac{ d^4k~\delta^+((k+p_2)^2-m^2) }
       {  (k^2-\lambda^2+i\epsilon)((k+p_4)^2-{m'}^2+i\epsilon)  }\times
\\ \nonumber
&&
{}~~~~~~~~~~\times\frac{1}{2}(P-k)^2\Theta ^+((P-k)^2)
\eeqn
where we have defined $\Theta ^+(v^2)=\Theta(v^2)\Theta(v_0)$.

\noindent
The corresponding integrals for the cuts $(i)$ and $(ii)$ are obtained
from the cut $(iii)$ respectively, with the substitutions:
\beqn
\frac{  \delta^+((k+p_2)^2-m^2)  }
       {  (k^2-\lambda^2)((k+p_4)^2-{m'}^2)  }
&\rightarrow&
\frac{  \delta^+((k+p_4)^2-{m'}^2)  }
       {  (k^2-\lambda^2)((k+p_2)^2-m^2)  }
\\ \nonumber
&\rightarrow&\frac{ \delta^+(k^2-\lambda^2) }
       {  ((k+p_2)^2-m^2)((k+p_4)^2-{m'}^2)  }
\eeqn
To simplify the computation of the integrals, we consider a non-relativistic
configuration of the external momenta, i.e. we take:
\beq\label{eq:nonrel}
p_1=p_3=(m_b,\vec{0})~~~{\rm and}~~~p_2=p_4=(m,\vec{0})
\eeq
The cuts $(i)$ and $(iii)$ have singularities in the forward limit, i.e. for
$p_1=p_3$ and $p_2=p_4$, and the configuration (\ref{eq:nonrel}) is a forward
one.
We regulate these singularities by taking a different
mass for the $\overline{u}$ quark in the initial state, $m_u=m$, and in the
final state, $m_u=m'$. After summing the cuts $(i)$ and $(iii)$, we are
allowed to take the forward scattering limit $m'\rightarrow m$.

\noindent
With our choice of the external momenta, $I_{\mu\nu}$ can be parametrized as:
\beq
I_{\mu\nu}~=~g_{\mu\nu}A+\frac{ p_{2\mu}p_{2\nu} }{m^2} B
\eeq
For the projections
\beq
X~=~g_{\mu\nu}I^{\mu\nu}~=~4A+B~~~~~{\rm and}~~~~~
Y~=~\frac{p_{2\mu}p_{2\nu}}{m^2}I^{\mu\nu}~=~A+B,~~~~
\eeq
we have, for the cut $(iii)$:
\beqn\label{eq:final1}
X_{(iii)}&=&\frac{-\pi}{\delta m}\int_0^{\overline{k}/m}
\frac{k^2 dk}{
2\sqrt{1+k^2}(\sqrt{1+k^2}-1)(\sqrt{1+k^2}-1+\lambda^2/2m^2)}~~~~
\\  \nonumber
&\times& [m-\delta m\sqrt{1+k^2}][(M+m)^2+m^2-2m(M+m)\sqrt{1+k^2}]
\\  \nonumber
 &=&\pi M^2[\frac{\pi m}{\lambda}+\frac{1}{8}\frac{M}{m}
                               -\frac{\pi m^2}{\delta m\lambda}+\ldots]
\eeqn
and
\beqn\label{eq:final2}
Y_{(iii)}&=&\frac{\pi}{\delta m}\int_0^{\overline{k}/m}
\frac{k^2 d
k}{2\sqrt{1+k^2}(\sqrt{1+k^2}-1)(\sqrt{1+k^2}-1+\lambda^2/2m^2)}~~~~
\\ \nonumber
&\times& \frac{1}{2}[(M+m)^2+m^2-2m(M+m)\sqrt{1+k^2}]
              [2m k^2+m-\delta m\sqrt{1+k^2}]
\\ \nonumber
&=&\pi M^2[-\frac{\pi}{2}\frac{m}{\lambda}-\frac{1}{16}\frac{M}{m}
           +\frac{1}{48}\frac{M^2}{m\delta m}
             +\frac{\pi}{2}\frac{ m^2}{\delta m\lambda}+\ldots]
\eeqn

\noindent
In eqs.(\ref{eq:final1}) and (\ref{eq:final2})
the shift $k\rightarrow k+p_2$ and the conversion to
an adimensional momentum $k/m$ have been done.

The dots '$\ldots$' indicate finite or at most logarithmic terms,
$\delta m=m-m'$, $M=m_b-m$ is the temporal component of $P$,
$P=(M,\vec{0})$,
and $\overline{k}=\overline{k}(M,m)$ is the upper limit for the gluon
3-momenta:
\beq\label{eq:theta}
\overline{k}=\sqrt{ \left(\frac{M+m}{2}\right)^2
        [1+\left(\frac{m}{M+m}\right)^2]-m^2 }
=\frac{M}{2}[1+\frac{m}{M}+\ldots]
\eeq
The analyticity argument is forced to disregard the $\theta$-function,
which is taken into account in eq.(\ref{eq:theta}).

\noindent
$X_{(iii)}$ and $Y_{(iii)}$ contain power singularities of the form:
\beq
\frac{1}{\lambda},~~~\frac{1}{\delta m}~~~{\rm and}~~~\frac{1}{m}
\eeq

\noindent
$X_{(i)}$ and $Y_{(i)}$ are computed from $X_{(iii)}$ and $Y_{(iii)}$
respectively by exchanging $m'$ with $m$.
Summing the contributions from the two cuts, we see that singularities
in $1/\lambda$ and $1/\delta m$ cancel, to give:
\beq\label{eq:partial}
X_{(i)}+X_{(iii)}~=~\pi M^2[ \frac{1}{4}\frac{M}{m}+\ldots ]
\eeq
and
\beq
Y_{(i)}+Y_{(iii)}~=~\pi M^2[-\frac{1}{48}\left(\frac{M}{m}\right)^2
-\frac{1}{8}\frac{M}{m}+\ldots]
\eeq

The computation of the cut $(ii)$ related to the real gluon emission diagram
is simpler because there is no the forward singularity (one can take
$m=m'$ from the beginning), and gives:
\beq
X_{(ii)}~=~\pi M^2\frac{1}{4}[-\frac{M}{m}+\ldots ]
\eeq
and
\beq
Y_{(ii)}=\pi M^2[\frac{1}{48}\left(\frac{M}{m}\right)^2+\frac{1}{8}\frac{M}{m}
+\ldots ]
\eeq
Summing the contributions from cuts $(i)$, $(ii)$ and $(iii)$, we see that
there
is a cancellation of the power singularities in $1/m^2$ and $1/m$.

The computation of $V_{\mu}^{(iii)}$ gives:
\beq
V_{\mu}^{(iii)}=\delta_{\mu 0}\pi M^2[~\frac{\pi}{2}\frac{m}{\delta m\lambda}
+\frac{1}{16}\frac{M}{m\delta m}+\ldots~]
\eeq
$V_{\mu}^{(i)}$ may be computed from $V_{\mu}^{(iii)}$ by means of the formula:
\beq
V_{\mu}^{(i)}~=~V_{\mu}^{(iii)}(m\rightarrow m')
            +\delta_{\mu 0}~\delta m ~ S^{(iii)} (m\rightarrow m')
\eeq
The sum of the cuts $(i)$ and $(iii)$ gives in this case:
\beq
m[V_{\mu}^{(i)}+V_{\mu}^{(iii)}]~=~\delta_{\mu 0}\pi M^2
                  [-\frac{1}{16}\frac{M}{m}+\ldots]
\eeq
while
\beq\label{eq:end}
mV_{\mu}^{(ii)}~=~\delta_{\mu 0}\pi M^2[~\frac{1}{16}\frac{M}{m}+\ldots]
\eeq
The sum of the cuts does not contain any more power singularities.
The computation of $W_{\mu}$ is completely
analogous to that of $V_{\mu}$ and we do not report the results.
For $S$ we have the expression:
\beq
S_{(iii)}~=~\pi M^2[~\frac{\pi}{2}\frac{1}{\delta m\lambda}
            +\frac{1}{4}\frac{1}{m\delta m}\log\frac{M}{m}+\ldots~]
\eeq
Using the symmetry relation $S_{(i)}=S_{(iii)}(m\rightarrow m')$,
we have for the partial sum:
\beq
m^2[~S_{(i)}+S_{(iii)}~]=\pi M^2[-\frac{1}{4}\log\frac{M}{m}+\ldots]
\eeq
Note that there is not any power singularity because of the $m^2$ at the
numerator coming from Dirac algebra.
$m^2 S_{(ii)}$ also is free from power singularities:
\beq
m^2 S_{(ii)}~=~\pi M^2[~\frac{1}{4}\log\frac{M}{\lambda}+\ldots~]
\eeq
In the total sum therefore we have only an infrared logarithmic singularity:
\beq
m^2[~S_{(i)}+S_{(ii)}+S_{(iii)}~]~=~
\pi M^2[~\frac{1}{4}\log\frac{m}{\lambda}+\ldots~]
\eeq
This infrared singularity, together with those neglected
in eqs.(\ref{eq:partial})-(\ref{eq:end}), cancels in the total sum of the
real and virtual gluon emission diagrams.

We consider now the diagrams in figs.2 and 3,
connecting  an external heavy line
with an external light line via a gluon.
The cuts $(i)$ on the gluon line have $1/m$ singularities, while the cuts
$(ii)$ on the light quark line have both $1/m$ and $1/\lambda$ singularities.
$1/m$ singularities cancel in the sum of the cuts $(i)$ and $(ii)$ of
each diagram separately, while $1/\lambda$ singularities cancel in the total
sum of the four cuts. The computation is analogous to the one described
and we do not give the details.

In conclusion, we have verified explicitly the cancellation of power
enhancements in the inclusive widths of non-spectator decays.

\vskip 1. truecm
\centerline{\bf Acknowledgment}
\vskip .5 truecm

\noindent
I wish to express particular thanks to Profs. G. Altarelli and
S. Petrarca for many useful discussions and for encouragement.

\newpage
\input FEYNMAN
\begin{displaymath}
\begin{picture}(40000,20000)(-10000,0)
\drawline\fermion[\E\REG](0,13000)[20000]
\drawarrow[\LDIR\ATTIP](2500,13000)
\drawarrow[\LDIR\ATTIP](10000,13000)
\drawarrow[\LDIR\ATTIP](17500,13000)
\drawline\fermion[\SW\REG](15000,13000)[14142]
\drawarrow[\LDIR\ATTIP](14000,12000)
\drawline\fermion[\W\REG](\pbackx,\pbacky)[5000]
\drawarrow[\LDIR\ATTIP](2500,3000)
\drawline\fermion[\W\REG](20000,3000)[5000]
\drawarrow[\LDIR\ATTIP](17500,3000)
\drawline\fermion[\NW\REG](\pbackx,\pbacky)[14142]
\drawarrow[\LDIR\ATTIP](6000,12000)
\drawline\fermion[\NE\REG](\pbackx,\pbacky)[7071]
\drawarrow[\LDIR\ATTIP](7500,15500)
\drawline\fermion[\SE\REG](\pbackx,\pbacky)[7071]
\drawline\gluon[\E\REG](7800,10200)[4]
\drawline\scalar[\S\REG](6500,18000)[8]
\drawline\scalar[\S\REG](10000,18000)[8]
\drawline\scalar[\S\REG](13500,18000)[8]
\put(6500,0){$(i)$}
\put(10000,0){$(ii)$}
\put(13500,0){$(iii)$}
\put(10000,-3000){\bf Fig.~1}
\put(0,14000){b}
\put(2500,14000){$p_1$}
\put(17500,14000){$p_3$}
\put(0,4000){$\overline{u}$}
\put(2500,4000){$p_2\rightarrow$}
\put(17500,4000){$p_4\rightarrow$}
\put(6500,14000){$P-k-l,~~d$}
\put(9000,19000){$l,~~c$}
\put(8800,11000){$k\rightarrow$}
\end{picture}
\end{displaymath}
\vskip .7 truecm
\begin{displaymath}
\begin{picture}(40000,20000)(-10000,0)
\drawline\fermion[\E\REG](0,13000)[20000]
\drawarrow[\LDIR\ATTIP](1250,13000)
\drawarrow[\LDIR\ATTIP](4750,13000)
\drawarrow[\LDIR\ATTIP](10000,13000)
\drawarrow[\LDIR\ATTIP](17500,13000)
\drawline\fermion[\SW\REG](15000,13000)[14142]
\drawarrow[\LDIR\ATTIP](14000,12000)
\drawline\fermion[\W\REG](\pbackx,\pbacky)[5000]
\drawarrow[\LDIR\ATTIP](2500,3000)
\drawline\fermion[\W\REG](20000,3000)[5000]
\drawarrow[\LDIR\ATTIP](17500,3000)
\drawline\fermion[\NW\REG](\pbackx,\pbacky)[14142]
\drawline\fermion[\NE\REG](\pbackx,\pbacky)[7071]
\drawarrow[\LDIR\ATTIP](7500,15500)
\drawline\fermion[\SE\REG](\pbackx,\pbacky)[7071]
\drawline\gluon[\SE\REG](2000,13000)[6]
\drawline\scalar[\S\REG](6500,18000)[8]
\drawline\scalar[\S\REG](11500,18000)[8]
\put(6500,0){$(i)$}
\put(11500,0){$(ii)$}
\put(10000,-3000){\bf Fig.~2}
\end{picture}
\end{displaymath}
\vskip .7 truecm
\begin{displaymath}
\begin{picture}(40000,20000)(-10000,0)
\drawline\fermion[\E\REG](0,13000)[20000]
\drawarrow[\LDIR\ATTIP](2500,13000)
\drawarrow[\LDIR\ATTIP](10000,13000)
\drawarrow[\LDIR\ATTIP](16250,13000)
\drawarrow[\LDIR\ATTIP](18750,13000)
\drawline\fermion[\SW\REG](15000,13000)[14142]
\drawline\fermion[\W\REG](\pbackx,\pbacky)[5000]
\drawarrow[\LDIR\ATTIP](2500,3000)
\drawline\fermion[\W\REG](20000,3000)[5000]
\drawarrow[\LDIR\ATTIP](17500,3000)
\drawline\fermion[\NW\REG](\pbackx,\pbacky)[14142]
\drawarrow[\LDIR\ATTIP](6000,12000)
\drawline\fermion[\NE\REG](\pbackx,\pbacky)[7071]
\drawarrow[\LDIR\ATTIP](7500,15500)
\drawline\fermion[\SE\REG](\pbackx,\pbacky)[7071]
\drawline\gluon[\SW\REG](18000,13000)[6]
\drawline\scalar[\S\REG](8500,18000)[8]
\drawline\scalar[\S\REG](13500,18000)[8]
\put(8500,0){$(ii)$}
\put(13500,0){$(i)$}
\put(10000,-3000){\bf Fig.~3}
\end{picture}
\end{displaymath}
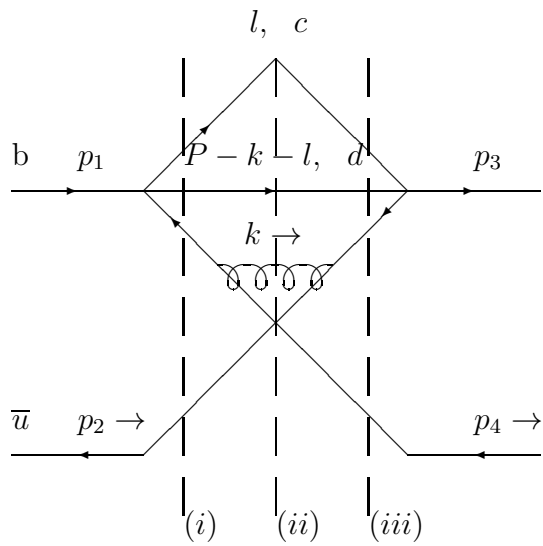
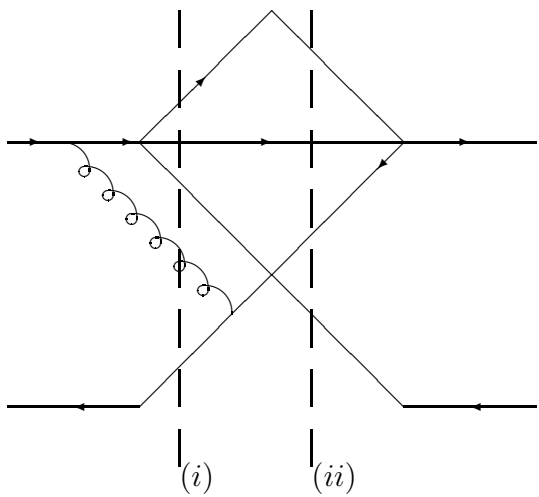
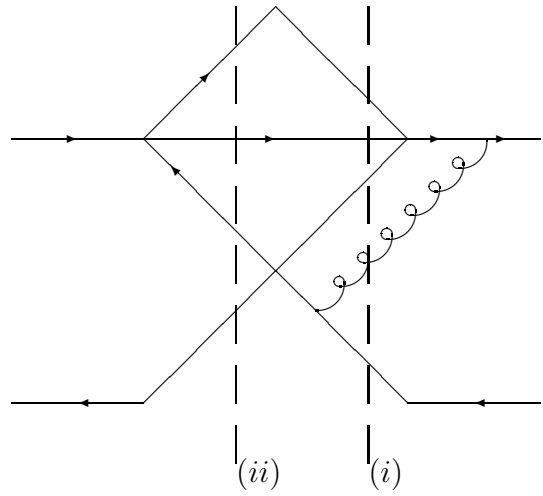
\end{document}